# A TCN-based Hybrid Forecasting Framework for Hours-ahead Utility-scale PV Forecasting

Yiyan Li, *Member, IEEE*, Lidong Song, *Student Member, IEEE*, Si Zhang, *Student Member, IEEE*, Laura Kraus, Taylor Adcox, Roger Willardson, Abhishek Komandur, and Ning Lu, *Fellow, IEEE*

*Abstract*—This paper presents a Temporal Convolutional Network (TCN) based hybrid PV forecasting framework for enhancing hours-ahead utility-scale PV forecasting. The hybrid framework consists of two forecasting models: a physics-based trend forecasting (TF) model and a data-driven cloud-event forecasting (CF) model. Three TCNs are integrated in the framework for: i) blending the inputs from different Numerical Weather Prediction sources for the TF model to achieve superior performance on forecasting hourly PV profiles, ii) capturing spatial-temporal correlations between detector sites and the target site in the CF model to achieve more accurate forecast of intra-hour PV power drops, and iii) reconciling TF and CF results to obtain coherent hours-ahead PV forecast with both hourly trends and intra-hour fluctuations well preserved. To automatically identify the most contributive neighboring sites for forming a detector network, a scenario-based correlation analysis method is developed, which significantly improves the capability of the CF model on capturing large power fluctuations caused by cloud movements. The framework is developed, tested, and validated using actual PV data collected from 95 PV farms in North Carolina. Simulation results show that the performance of 6 hours ahead PV power forecasting is improved by approximately 20% compared with state-of-the-art methods.

*Index Terms*—Detector network, neighbor selection, NWP blending, physics-based model, short-term PV forecast, spatial-temporal forecasting, temporal convolutional network.

## NOMENCLATURE

*Scalar*
| | |
|---|---|
| $d$ | Dilation rate |
| $D$ | Number of days in the historical data |
| $E_{bias}$ | Bias in the physics-based model |
| $h$ | Index of the historical days |
| $K$ | Filter size |
| $M$ | Length of the daily irradiance profile |
| $N$ | Number of PV sites |
| $N_{stack}$ | Number of module stacks in the TCN model |
| $p_{real}^t$ | Actual power output from field measurements |
| $p_{simu}^t$ | Power output of the physics-based model at time $t$ |
| $P_{cc}$ | Pearson Correlation Coefficient |
| $P_{cc.\max}$ | $P_{cc}$ value with the optimal time shift, $\Delta t_{\max}$ |
| $R_{field}$ | Receptive field of the TCN model |
| $S_h$ | Index of the correlation scenarios on $h^{\text{th}}$ day |
| $t$ | Index of the time series data |
| $\Delta t$ | Time shift between two time series |
| $\Delta t_{\max}$ | Optimal time shift that leads to $P_{cc.\max}$ |
| $T_{shift}$ | Threshold of time-lagged correlation analysis |
| $T_{thre}$ | Threshold of $\Delta t_{\max}$ to determine successful detection |
| $\Delta x$ | Irradiance change threshold to detect cloud event |
| $\varphi$ | Successful detection rate |
| $\varphi_{\max}$ | Maximum successful detection rate |

*Vector/Matrix*
| | |
|---|---|
| $\mathbf{F}$ | Filter in TCN, $\mathbf{F} = [f_0, f_1, \ldots, f_{K-1}]$ |
| $\mathbf{X}$ | Normalized irradiance vector, $\mathbf{X} = [x^1, x^2, \ldots, x^M]$ |
| $\mathbf{X}_T$ | $\mathbf{X}$ of the target site, $\mathbf{X}_T = [x_T^1, x_T^2, \ldots, x_T^M]$ |
| $\mathbf{X}_D$ | $\mathbf{X}$ of the detector site, $\mathbf{X}_D = [x_D^1, x_D^2, \ldots, x_D^M]$ |
| $\hat{\mathbf{X}}_T$ | Differential vector of $\mathbf{X}_T$, $\hat{\mathbf{X}}_T = [\hat{x}_T^1, \hat{x}_T^2, \ldots, \hat{x}_T^M]$ |
| $\hat{\mathbf{X}}_D$ | Differential vector of $\mathbf{X}_D$, $\hat{\mathbf{X}}_D = [\hat{x}_D^1, \hat{x}_D^2, \ldots, \hat{x}_D^M]$ |
| $\mathcal{J}$ | Vector of the neighboring sites |
| $\mathcal{F}$ | Vector of the detector network |
| $\mathcal{F}_{opt}$ | Vector of the selected optimal detector network |
| $\mathcal{T}$ | Matrix of $\Delta t_{\max}$, $D \times (N-1)$ |
| $\mathcal{P}$ | Matrix of $P_{cc.\max}$ |
| $\mathbf{\Phi}$ | Vector of $\varphi$, $1 \times (N-1)$ |

*Functions*
| | |
|---|---|
| $\mathcal{G}$ | Dilated convolution operator |
| $\mathcal{I}_A$ | Indicator function with condition set $A$ |

## I. INTRODUCTION

THE stochasticity of cloud movements is the major cause for large PV forecasting errors, especially in the hours-ahead time horizon. Existing PV forecasting methods can be divided into *physics-based models* and *data-driven models* [1]-[3]. Physics-based models use irradiance forecasts (usually obtained from Numerical Weather Prediction (NWP), satellite images, total sky imagers, etc.) as inputs to forecast PV power outputs. Once a physics-based model is well-calibrated, its forecasting accuracy and resolution will completely rely on the forecasting accuracy of its input data. Meanwhile, because the physics-based model requires detailed parameters of PV modules/inverters, it requires considerable efforts for maintaining model accuracy. Therefore, the physics-based

This material is based upon work supported by the U.S. Department of Energy's Office of Energy Efficiency and Renewable Energy (EERE) under the Solar Energy Technologies Office. Award Number: DE-EE0008770. Yiyan Li, Lidong Song, Si Zhang and Ning Lu are with the Electrical & Computer Engineering Department, Future Renewable Energy Delivery and Management (FREEDM) Systems Center, North Carolina State University, Raleigh, NC 27606 USA. (yli257@ncsu.edu, lsong4@ncsu.edu, szhang56@ncsu.edu, nlu2@ncsu.edu ). Laura Kraus, Taylor Adcox, Roger Willardson and Abhishek Komandur are with Strata Clean Energy, Durham, NC 27701 USA. (lkraus@stratacleanenergy.com, tadcox@stratacleanenergy.com, rwillardson@stratacleanenergy.com, akomandur@stratacleanenergy.com)

modeling approach is usually used for modeling MW-level PV farms and rarely applied to residential rooftop PV systems.

Data-driven models do not require parameters from PV system components. Methods, such as regression or machine learning based models, can be used for predicting the future PV power output from historical data assuming that the future follows similar patterns as the past. Because weather conditions can change rapidly in a day, the data-driven model is usually used for hours-ahead forecasting while the physics-based methods for day-ahead PV forecasting. The two methods are further compared in Table I.

TABLE I
QUALITATIVE COMPARISON BETWEEN THE PHYSICS-BASED MODEL AND DATA-DRIVEN MODEL

|  | Physics-based model | Data-driven model |
|---|---|---|
| Historical data requirement | Low | High |
| NWP requirement | High | Low |
| Physical parameter requirement | High | Low |
| Model adaptability | Low | High |
| Forecasting resolution | Same as NWP | Flexible |
| Effective forecasting horizon | Days-ahead | Hours-ahead |

Recently, leveraging spatial-temporal correlations between neighboring sites for capturing cloud movements to improve short-term PV forecasting accuracy, as an emerging branch of data-driven method, has drawn increasing attention. By analyzing power outputs of a group of PV sites or weather data from adjacent meteorological stations, patterns of cloud movements can be accounted for implicitly or explicitly. Regression models, such as AR, ARX, NARX, are firstly introduced in [4]-[7] to build the mapping between neighboring sites and the target site. In [8], Yang *et al.* achieve spatial-temporal PV forecasting using time-forward kriging. Compressive sensing is used by Tascikaraoglu *et al.* in [9] to extract the spatial-temporal correlations between a target meteorological station and its neigboring stations to improve the short-term forecasting accuracy. In [10], Zhang *et al.* use Bayesian network to forecast PV generation based on spatial-temporal analysis results. Recently, deep-learning receives increasing attention because of its powerful nonlinear learning ability. In [11], Liu *et al.* combine CNN and GRU to extract the spatial-temporal information from multi-dimensional time series inputs from adjacent sites. Similarly, in [12], Wang *et al.* implement CNN to extract the spatial features across multiple sites, and then use LSTM to achieve the forecasting of the target site. In [13], Jeong *et al.* organize the multi-site time series data into spatial-temporal matrix, and then use CNN to achieve feature extraction and forecasting. In [14], Venugopal *et al.* compare different CNN structures in extracting features from heterogeneous data sources, as well as their short-term PV forecasting performances. Compared with the forecasting methods using total sky imagers or satellite images, spatial-temporal forecasting methods are usually cheaper and easier to implement while achieving comparable performance without requiring extensive hardware or external data supports.

However, we identify two research gaps in the state-of-the-art methods. *First, there lacks of a deep fusion approach for seamlessly integrating physics-based and data-driven models into a hybrid PV forecasting framework to leverage both of their advantages.* As shown in Table I, the physics-based model can achieve stable hourly irradiance-power conversion for a much longer forecasting horizon than the data-driven model. However, it cannot effectively capture intra-hour power fluctuations caused by cloud movements. The data-driven model can forecast intra-hour PV fluctuations with more accuracy by capturing cloud movements from near real-time measurements. However, the forecasting accuracy and stability of data-driven model decrease rapidly when the forecasting horizon is longer than a few hours. Thus, the deep fusion between the physics-based model and the data-driven model has the potential to achieve both forecasting stability and finer granularity. To the best of the author's knowledge, such studies are rarely seen in existing literatures.

*Second, there lacks an automated, objective detector site selection mechanism.* Selecting the most contributive neighbors, i.e., the "detector" sites, for the "target" site among all its neighbors is crucial for improving forecasting accuracy and reducing model complexity in spatial-temporal forecasting models. Conventional detector site selection methods ignore the sequence of cloud events (i.e. which site sees the cloud event first and which next). In addition, existing methods are pairwise-correlation-based and cannot exploit the benefit of "collaboration", where a certain combination of neighboring sites can formulate a more powerful detector network. Lastly, conventional methods usually involve using manually-defined correlation thresholds for eliminating "uncorrelated" neighbors, making the process highly subjective.

Therefore, in this paper, we develop a Temporal Convolutional Network (TCN) based hybrid PV forecasting framework for enhancing hours-ahead utility-scale PV forecasting. The hybrid framework consists of two forecasting models: a physics-based trend forecasting (TF) model and a data-driven cloud-event forecasting (CF) model. Three TCNs are integrated in the framework: in the TF model, the first TCN is used for blending TF input data from different NWP sources into a coherent input data set that are then fed into the inverter-level physics-based model to obtain more accurate trend forecasting results. In the CF model, the second TCN is used for hours-ahead, intra-hour PV power forecasting, where predicting power fluctuations caused by large cloud events is crucial. Using measurements from selected detector sites as inputs, the second TCN can generate more accurate intra-hour PV forecasts with minute-level granularity. Then, the third TCN is used to reconcile the TF and CF results. Using a sequence-to-sequence model for achieving temporal consistency between the two time series forecasts, we can further improve the short-term PV forecasting accuracy by capturing both the hourly trends and intra-hour PV fluctuations.

In summary, the contributions of this paper are two-folds. *First*, the proposed hybrid PV forecasting framework seamlessly combines the advantages of the physics-based model and the data-driven model so that both the hourly trend and the intra-hour fluctuations can be adequately predicted. *Second*, we develop a scenario-based, automated detector site selection algorithm to identify the most contributive neighbors for a target site. One distinct advantage of the detector selection algorithm is that we consider the temporal leading/lagging patterns and the collaborative effect between sites. The proposed algorithm is automated and objective so neither domain expertise nor human supervision is required.

The rest of the paper is organized as follows: Section II introduces the proposed hybrid PV forecasting framework. Section III demonstrates the simulation results. Section IV concludes this paper.

## II. METHODOLOGY

The proposed hybrid PV forecasting framework for hours-ahead PV forecasting is illustrated in Fig.1. The framework consists of two forecasting models: TF and CF. The TF model is a physics-based model for predicting hourly PV power outputs over a time horizon of up to a week. The CF model is a data-driven model for predicting intra-hour PV output fluctuations at 5-minute resolution for the next 1 to 6 hours.

The input of the TF model is the hourly NWP data. To improve the TF model performance, a TCN-based NWP data blender (TCN #1, detailed in Section II.B) is developed to fuse the data from different NWP sources with an objective of maximizing the forecast accuracy of irradiance inputs. Then, the predicted irradiance from TCN is converted into hourly PV forecasts by the physics-based model.

The inputs of the CF model are the normalized historical irradiance data from the PV sites in the detector network. To select the most-correlated neighbors to form the detector network, we develop a *scenario-based neighbor selection* algorithm for automatically identifying an effective detector network for the target site without human supervision (detailed in Section II.C). First, at each time step, the irradiance data collected at all PV sites will be coded into a two-dimensional (2D) matrix based on their geographical locations. Then, the historical data of the target site together with the detector sites will be fed into the TCN-based CF model (TCN #2) to extract the spatial-temporal information to produce the intra-hour PV forecasting.

To perform forecast reconciliation, forecasting results from both TF and CF models will be fed to a TCN (TCN #3, detailed in Section II.D) to eliminate the inconsistencies (i.e., different data resolution and magnitude discrepancies) between the two time-series profiles. The resultant hours-ahead PV forecast with 5-minute resolution captures both the hourly trend and intra-hour fluctuations by combining the advantages of two independent forecasting models, namely the physics-based model and the data-driven model.

In practice, the TF model can be updated every 6 hours (same as the NWP data update cycle) while the CF model can be updated every 5-minute or longer depending on the communication network setup between the control center and each PV site.

### A. Temporal Convolutional Network

TCN is a fully convolutional-based network structure [15], as shown in Fig. 2. Each convolutional layer needs to have the same length as the input layer. To meet this requirement, zero padding is applied to solve the dimension reduction issue caused by the convolution operation (see the dashed blocks in Fig. 2). After a few convolutional layers, the features of the input time series are extracted and compressed into the output layer, which can be further used for forecasting purpose. Compared with sequential networks such as RNN and LSTM, such a purely convolutional structure of TCN can be highly paralleled in model training and therefore has better training efficiency [15]. TCN has been successfully used to solve the time series forecasting problems, such as PM2.5 forecasting [16] and load forecasting [17], etc. However, it has not yet been introduced for conducting PV forecasting.

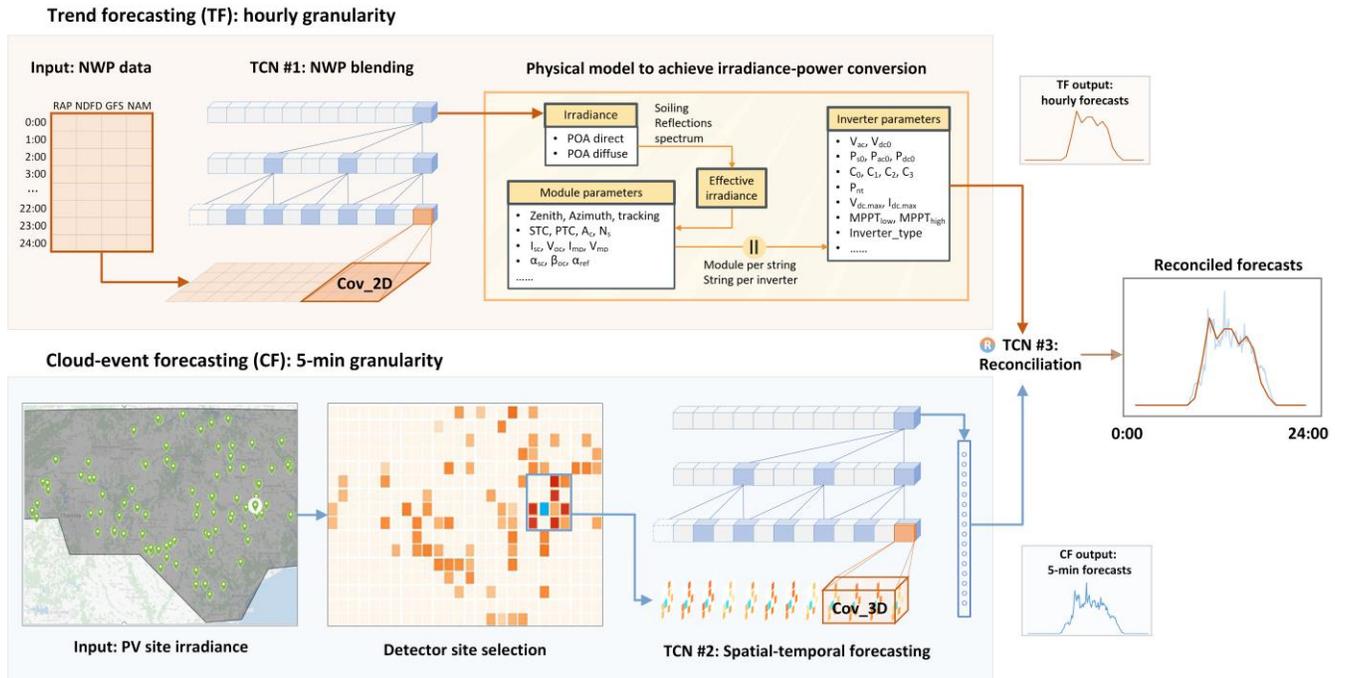

Fig. 1. Flowchart of the proposed hybrid PV forecasting framework

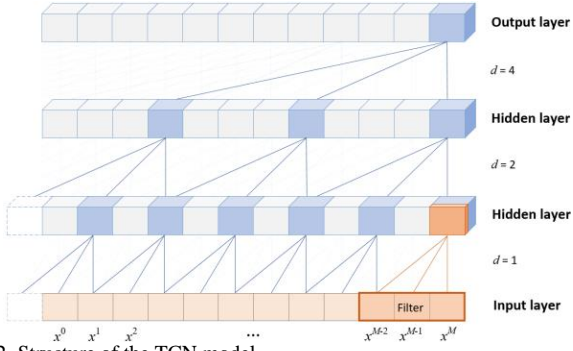

Fig. 2. Structure of the TCN model

The most significant feature of TCN is the *dilated convolution*. Assume the input time series is $\mathbf{X}=[x^0, x^1, \ldots, x^M]$, and we use a filter $\mathbf{F}=[f_0, f_1, \ldots, f_{K-1}]$ to conduct convolution. Then, the dilated convolution $\mathcal{G}(\cdot)$ for the element $x^m$ in $\mathbf{X}$ can be calculated by

$$\mathcal{G}(x^m) = \sum_{i=0}^{K-1} f_i \cdot x^{m-d \cdot i} \quad (1)$$

where $d$ is the dilation rate, and $m$-$d\cdot i$ indexes to the past historical data before $x^m$. When $d =1$, (1) reduces to a regular convolution operation (e.g. the first layer in Fig. 2). When $d$ is larger than 1 (e.g. the second and third layers in Fig. 2), the filter will skip $(d-1)/d$ of the elements in the previous hidden layer and only focus on the remaining $1/d$. In this way, the whole network will have a large *receptive field* that increases quickly with the number of layers with limited model complexity. The receptive field of TCN model can be calculated by

$$R_{field} = 1 + 2 \cdot (K-1) \cdot N_{stack} \cdot \sum_i d_i \quad (2)$$

According to (2), we have 3 ways to increase the receptive field: using larger filter size $K$, larger dilation rate $d$, and increasing the network depth $N_{stack}$.

*Causal convolution* is another feature of the TCN model. As shown in Fig.2, each convolution layer only extracts information from the past historical data. In other words, there is no "information leakage" from the future. This structure is particularly suitable in solving forecasting problems where the future information is unavailable. To further improve the model performance, residual connections [18] can be used to achieve identical mappings.

### B. TCN-based NWP Data Blending

In practice, inverter-level physics-based models are commonly used to convert the irradiance (obtained from the NWP data) to power based on the inverter used at each PV site. However, the NWP data can come from different sources, for example, High-Resolution Rapid Refresh (HRRR), Global Forecast System (GFS), National Digital Forecast Database (NDFD), Rapid Refresh (RAP) and North American Mesoscale (NAM). As shown in Table II, because different NWP data sources have different forecasting features and data granularity, when using different NWP data as inputs, discrepancy in TF model arises.

This inspires us to develop a data blending model to merge the NWP data from different sources together with an objective to offset modeling deficiencies and improve the overall forecasting performance. Thus, we developed a TCN-base NWP data blending model (See TCN #1 in Fig.1) to achieve the seq2seq mapping from NWPs to the actual irradiance. The inputs are time-series data from different NWP data sources, and the output is the time series of the field measurement irradiance.

TABLE II
FEATURES OF DIFFERENT NWP DATA SOURCES AND THEIR FORECASTING PERFORMANCE

|  | HRRR | GFS | NAM | NDFD | RAP |
|---|---|---|---|---|---|
| Spatial resolution | 3km | 28-44km | 12km | 2.5km | 13km |
| Dara granularity | 1h | 3h | 1-3h | 1h | 1h |
| Forecasting horizon | 15h | 16days | 4days | 36h | 1day |
| Forecasting bias | 23.74 | 6.89 | 14.99 | 1.31 | 16.11 |
| Forecasting RMSE | 76.14 | 82.37 | 80.30 | 61.72 | 68.81 |

### C. Scenario-based Detector Sites Selection Algorithm

The main challenge in short-term PV forecast is to forecast, in a few hours-ahead, the large power drops caused by cloud movements. Thus, the objective of the detector sites selection algorithm is to select an optimal group of detector sites so that using the solar irradiance data of those detector sites as inputs, large PV output drops at the target site can be more accurately predicted. Note that, in this paper, we focus our analysis on detecting cloud events instead of the entire irradiance time series.

As shown in Fig.3(a), the irradiance data used in our study is in 5-minute granularity. Thus, there are 288 data points in a day (i.e., $M = 288$). Normalize all the daily irradiance profiles for all PV sites. Let the target site irradiance profile be $\mathbf{X}_T = [x_T^1, x_T^2, \ldots, x_T^M]$ and the detector site be $\mathbf{X}_D = [x_D^1, x_D^2, \ldots, x_D^M]$.

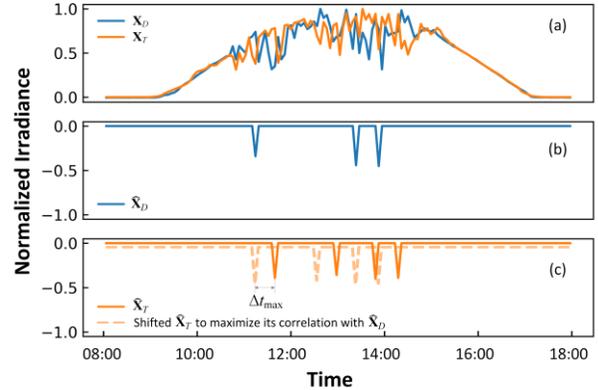

Fig. 3. Example of the event-based correlation analysis method. (a) Original daily irradiance series of the target site and the detector site. (b) Cloud events at the detector site. (c) Cloud events and the maximum correlation time shift at the target site.

Define a cloud event as the irradiance drop greater than threshold $\Delta x$ during two consecutive time intervals

$$x_t - x_{t-1} \leq -\Delta x \quad (3)$$

Note that $\Delta x$ is dependent on data granularity. In this paper, we define $\Delta x = 0.3$. This is because we want to capture the cloud event that can cause a power drop greater than 30% of the rated power in 5 minutes.

To extract the cloud event for the target site and the detector site, $\hat{\mathbf{X}}_T$ and $\hat{\mathbf{X}}_D$, respectively, we have

$$\hat{\mathbf{X}}_T = [\hat{x}_T^1, \hat{x}_T^2, ..., \hat{x}_T^{M-1}] \qquad (4)$$
$$\hat{\mathbf{X}}_D = [\hat{x}_D^1, \hat{x}_D^2, ..., \hat{x}_D^{M-1}]$$

$$\hat{x}_T^i, \hat{x}_D^i = \begin{cases} 0, & \hat{x}_T^i, \hat{x}_D^i > -\Delta x \\ \hat{x}_T^i, \hat{x}_D^i, & \hat{x}_T^i, \hat{x}_D^i \leq -\Delta x \end{cases} \qquad (5)$$

From the irradiance profiles of the two PV sites shown in Fig. 3(a), we can extract the cloud event series of the detector and target sites as shown by the solid lines in Fig. 3(b) and 3(c).

Next, a time-lagged correlation analysis [19] is conducted to find the optimal time shift $\Delta t_{max}$ between the target and the detector series so that the Pearson Correlation Coefficient [20] between the two sites, $P_{cc}$, can be maximized. The problem is formulated as

$$\Delta t_{max} = \underset{-T_{shift} \leq \Delta t \leq T_{shift}}{\arg\max} P_{cc}(\hat{\mathbf{X}}_T[\Delta t: M + \Delta t - 1], \hat{\mathbf{X}}_D) \qquad (6)$$

The indexed values out of range [1: $M$-1] are padded by 0. In (6), we select $T_{shift}$ = 8h. Note that $T_{shift}$ is a constant threshold of $\Delta t$ that guarantees the correlation calculation is within the same-day horizon. As shown in Fig. 3(c), after $\hat{\mathbf{X}}_T$ is shifted by $\Delta t_{max}$ (the dotted line), it is well aligned with $\hat{\mathbf{X}}_D$.

After calculating the $P_{cc}$ between the target site and detector sites, existing methods [5][10] select detectors with the highest $P_{cc}$ values under the assumption that a site with high $P_{cc}$ indicates higher contribution to the forecasting accuracy of the target site. However, such methods have three drawbacks:

- The approach ignores the temporal correlation crucial to detect cloud events. Only sites with a leading correlation with the target site will contribute to the target site forecasting results. This is because only when the cloud passes the detector site earlier than the target site ($\Delta t_{max} > 0$), can the information be used to to forecast the target site cloud events. A detector site having a high $P_{cc}$ but with lagging correlation ($\Delta t_{max} \leq 0$) cannot foresee the upcoming cloud event on the target site.
- The selection of the $P_{cc}$ threshold is subjective and empirical. In practice, different neighbor sites have different $P_{cc}$ with the target site. A $P_{cc}$ threshold is therefore needed to eliminate low-correlated neighbors. However, because the $P_{cc}$ value can vary significantly for different cases, it is difficult to manually select optimal thresholds or optimal number of detector sites.
- When detector sites are selected pair-by-pair using $P_{cc}$, the synergetic effect among detectors in a network setting cannot be properly accounted for.

Therefore, in this paper, we propose a scenario-based detector sites selection algorithm to overcome those drawbacks.

First, all the 6 possible correlation scenarios between the target and the neighbor sites are summarized in Table III, based on their daily cloud conditions and $\Delta t_{max}$. If there is at least one cloud event during the day, this day is defined as cloudy. Otherwise, this day is defined as clear sky. Note that Scenario 1 is not considered because our focus is to detect the cloud events that will significantly impact the PV output.

Among the six scenarios, Scenario 5 is defined as "successful detection". This means that a cloud event occurs earlier in the detector site than in the target site with a leading time $\Delta t_{max} \in (0, T_{thre}]$. Note that $T_{thre}$ equals to the forecasting horizon in the CF model to pick out the most contributive neighbors (e.g., in a 2-hour ahead forecasting scenario, $T_{thre}$=2).

TABLE III
DIFFERENT CORRELATION SCENARIOS

| Scenario No. | Detector site | Target site | $\Delta t$ | Definition |
|---|---|---|---|---|
| 1 | Clear sky | Clear sky | \ | Ignored |
| 2 | Cloudy | Clear sky | \ | Wrongly detect |
| 3 | Clear sky | Cloudy | \ | Fails to detect |
| 4 | Cloudy | Cloudy | $\Delta t_{max} \leq 0$ | Fails to detect |
| **5** | **Cloudy** | **Cloudy** | $\mathbf{0 < \Delta t_{max} \leq T_{thre}}$ | **Successful detection** |
| 6 | Cloudy | Cloudy | $T_{thre} < \Delta t_{max}$ | Irrelevant |

Based on Table III, we further define the successful detection rate for a detector site as

$$\varphi = \frac{\sum_{h=1}^{D} \mathcal{I}_{[5]}(S_h)}{\sum_{h=1}^{D} \mathcal{I}_{[2,3,4,5,6]}(S_h)} \qquad (7)$$

where $\mathcal{I}_A(x)$ is the indicator function that

$$\mathcal{I}_A(x) = \begin{cases} 1, & if \ x \in A \\ 0, & if \ x \notin A \end{cases} \qquad (8)$$

For a given detector-target site pair, $\varphi \in [0, 1]$ measures the successful detection rate. A larger $\varphi$ means the neighbor site has a statistically significant leading correlation pattern with the target site, indicating that the wind direction has a higher chance to blow from the detector site to the target site. Thus, $\varphi$ is an indicator that reflects localized weather patterns dominated by geographical characteristics.

To maximize $\varphi$, a detector network containing multiple neighbor sites is needed to have a better chance to forecast the upcoming cloud events for the target site. When calculating $\varphi$ of a detector network, for each historical day we select the neighbor with the largest $\Delta t_{max}$ as a representative because it has the best prediction ability. However, selecting an optimal subset among all candidate neighbors is a typical NP-hard problem. Therefore, in this paper, we design a greedy-searching algorithm to find a near-optimal solution. Key steps of the algorithm are summarized as follows:

*Step 1: Data Preparation.* Calculate the maximum time-lagged correlation coefficients, $P_{cc.max}$, and the corresponding $\Delta t_{max}$ between each candidate neighbor site and the target site based on historical cloud event data.

*Step 2: Forming the Detector Network.* Add each neighboring site successively to the detector network in descending order base on yearly averaged $P_{cc.max}$. Every time a new neighbor is included, recalculate $\varphi$ for the detector network. After all potential neighbors are added into the detector network, the detector network with the maximum $\varphi$ value will be selected.

*Step 3: Network Refinement.* The goal of refinement is to remove "bad" neighbors by successively removing the neighbors in the detector network to see if $\varphi$ can be further improved. If $\varphi$ is improved after a "bad" neighbor is removed, go back to step 2) and redo the detector network forming with the bad sites removed. If $\varphi$ cannot be improved, algorithm stops.

The pseudocode of the algorithm is shown in Algorithm 1. After selecting the detector network, we will further put the historical data of both the target site and the detector network into the TCN model to extract their spatial-temporal correlations, shown as TCN #2 in Fig.1. Results will be discussed in Section III.B.

---

**Algorithm 1**: Scenario-based detector site selection algorithm

**Input**: Target site $i$, neighboring sites $\mathcal{J}_{1\times(N-1)}$, cloud event sets of each site
**Output**: Selected detector network $\mathcal{F}_{opt}$ for the target site
**Initialization**: $\varphi_{max}=0$, $\mathcal{F}=[]$, $\mathcal{T}_{D\times(N-1)}=0$, $\mathcal{P}_{D\times(N-1)}=0$, flag=1

  # step 1): data preparation
1:  **for** $d = [1,2,…,D]$ **do**
2:    **for** $j = [1,2,…,N]$ and $j \neq i$ **do**
3:      calculate $\mathcal{T}(d, j) = \Delta t_{max}$ between site $i$ and $j$ according to (4)
4:      calculate $\mathcal{P}(d, j) = P_{cc.max}$
5:  **sort** $\mathcal{J}$ in descending order according to the average values of $\mathcal{P}$
6:  **while**(flag) **do**
7:    flag = 0
  # step 2): detector network formulation
8:    **for** $k = [1,2,…,N-1]$ **do**
9:      add site $\mathcal{J}[k]$ to $\mathcal{F}$
10:     calculate $\varphi$ of $\mathcal{F}$
11:     **if** $\varphi > \varphi_{max}$ **do**
12:       $\varphi_{max} = \varphi$, $\mathcal{F}_{opt} = \mathcal{F}$
  # step 3): detector network refinement
13:   **for** $k = [1,2,…, \text{length}(\mathcal{F}_{opt})]$ **do**
14:     calculate $\varphi$ of $\mathcal{F}_{opt}$ without $\mathcal{F}_{opt}[k]$
15:     **if** $\varphi > \varphi_{max}$ **do**
16:       remove $\mathcal{F}_{opt}[k]$ from $\mathcal{F}_{opt}$
17:       $\varphi_{max} = \varphi$
18:       flag = 1

---

*D. TCN-based Forecasting results reconciliation*

After obtaining the forecasting results under both TF and CF models, it is important to reconcile the results for two reasons. *First*, because different models are used for producing TF and CF forecasts, inconsistency in results are inevitable. For example, the hourly averages of the 5-min CF forecasts may deviate from the TF hourly forecasts. The inconsistency need to be reconciled so that the system operators know which value to use for future hours. *Second*, on the one hand, using weather data as inputs and taking a physics-based modeling approach, the TF model captures the PV average power output over a longer forecasting horizon well. On the other hand, the data-driven, TCN-based CF model captures intra-hour cloud events well. Thus, reconciliation can achieve mutual-benefit and improve the overall forecasting performance.

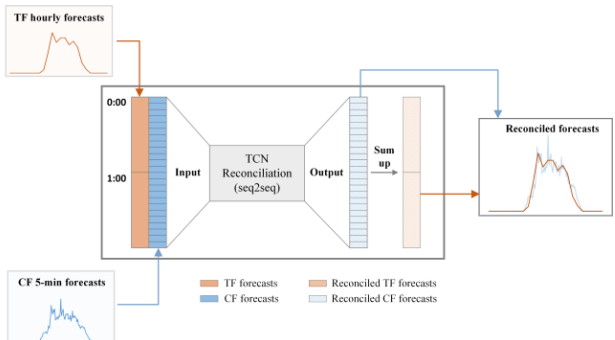

Fig. 4. Flowchart of the reconciliation process.

Reconciliation was introduced in the 2010s to solve the inconsistency in hierarchical forecasting problems. In the literature, analytical and machine-learning based reconciliation methods were proposed from both spatial and temporal aspects [21]-[26]. In this paper, we propose to use the sequence-to-sequence (seq2seq) mode of TCN for reconciliation. As shown in Fig. 4, the TF and CF forecasts are firstly aligned by time so that each hourly TF forecasting point is paired with 12 CF forecasting points (5-min granularity). This serves as the input of the reconciliation model. The output is compared with the actual 5-min power output profile of the target PV farm during the same time period. The reconciliation model is trained to fine-tune the CF forecasts under the guidance of the TF forecasts. Once the model is trained, the reconciled forecasting results are converted to two consistent sets of 5-minute and hourly PV outputs.

## III. CASE STUDY

In this paper, 5-minute field data collected from 95 utility-scale PV farms in North Carolina from 1/1/2020 to 11/30/2020 are used to develop and verify the performance of the proposed algorithm. Sizes of the PV farms range from 0.4MW to 26.2MW with locations shown in Fig. 5. The available measurements of each site are summarized in Table IV. For each PV site, irradiance measured by the pyranometer is selected to derive cloud movements because pyranometers are placed horizontally, making the measurement more comparable across different PV sites [27]. Missing points in pyranometer measurements (approximately 87% completeness) are patched using the correlations between pyranometer measurements and other irradiance-related measurements (e.g. inverter-level power output). As shown in Fig. 6, all measurements are normalized by their maximum value.

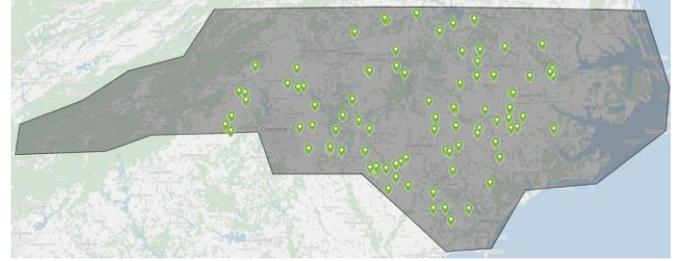

Fig. 5. PV farm locations, North Carolina, USA.

TABLE IV
AVAILABLE MEASUREMENTS AT EACH PV SITE

| Available measurements | Units |
|---|---|
| Energy export | kWh |
| Inverter-level power output | kW |
| Irradiance measured by reference cell | $W/m^2$ |
| Irradiance measured by pyranometer | $W/m^2$ |
| Ambient temperature | °C |
| Module temperature | °C |

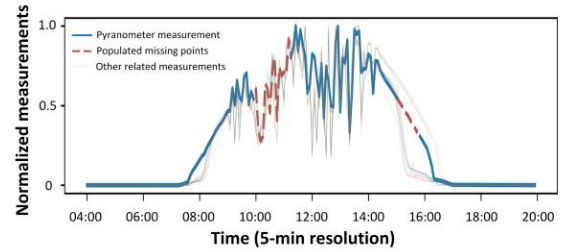

Fig. 6. An illustration of the missing data and the patching results.

## A. TF model: TCN-based NWP Data Blending

Five sources of NWP data (See Table II) from May to October in 2020 are fed into the physics-based model built by PVlib [28] to generate the TF forecasts. Historical data from May to September with hourly granularity are used to train the TCN model and test its performance in October. The TCN model configuration is shown in Table IX in the appendix.

First, we compare the TF forecaster performance with and without the TCN blending model. Then, we compare TCN with five commonly used benchmarking blending methods: Linear Regression (LR), Random Forest (RF), Support Vector Regression (SVR), Multilayer Perceptron (MLP) and LSTM. As shown in Table V, TCN reduces the TF forecasting bias to 0.47 and RMSE (Root Mean Squared Error) to 43.17, significantly outperform the NWP models without input blending and the five benchmarking methods. The forecasting results of Leo-B2 inverter from Oct.1 to Oct.7 are shown in Fig. 7 to illustrate the time-series forecasting results.

TABLE V
PERFORMANCE COMPARISON OF DIFFERENCE BLENDING METHODS

| Blending methods | Forecasting RMSE | Forecasting bias |
|---|---|---|
| LR | 52.39 | -9.81 |
| RF | 50.57 | -13.79 |
| SVR | 50.96 | 4.97 |
| MLP | 54.42 | -2.33 |
| LSTM | 48.01 | 1.52 |
| **TCN** | **43.17** | **0.47** |

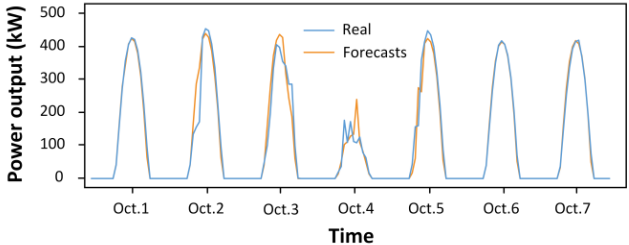

Fig. 7. TF forecasting results from Oct.1 to Oct.7.

## B. CF model: TCN-based spatial-temporal forecasting

In the CF model, we firstly apply the proposed detector site selection algorithm to automatically identify the most contributive neighbors for the target site. The cloud event detection threshold is $\Delta x = 0.3$, and the time-lagging threshold is $T_{thre} = 1h$ (for one hour-ahead forecasting as an example).

As shown in Fig. 8, Leo and Mars are used to demonstrate the proposed scenario-based detector site selection algorithm. At both sites, initially, the successful detection rate $\varphi$ will increase when the number of selected neighbors increases. However, when 7 neighbors are selected, $\varphi_{max}$ is reached. This shows that using information from a certain combination of leading-event neighbors helps detecting an upcoming cloud event, while including irrelevant neighbors can pollute the detection accuracy.

After the detector sites are selected, historical data of those sites are used train the TCN model to extract their spatial-temporal correlations for 1-hour ahead forecasting. We compare the TCN model with three state-of-the-art deep learning models that are commonly used for spatial-temporal PV forecasting: Convolutional Neural Network combined with Long-Short Term Memory (CNN-LSTM) [31], VGG-8 (Visual Geometry Group model with 8 layers) [32][33], and GARNN (Graph Attention Recurrent Neural Network) [34][35]. To make results comparable, similar model complexity is used, as shown in the Appendix.

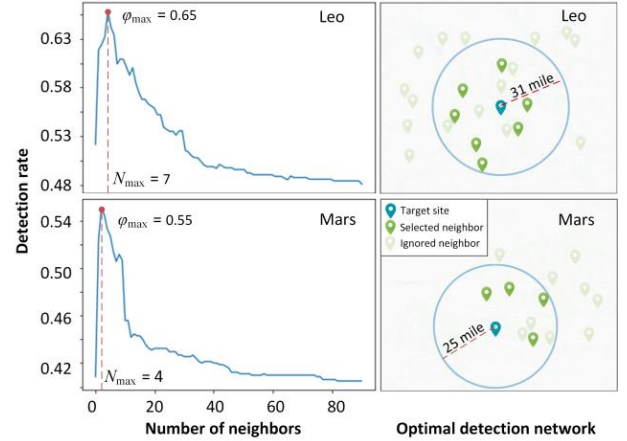

Fig. 8. Examples of the proposed detector site selection algorithm on site Leo and site Mars.

The 1-year historical data are split into training (70%), validation (10%), and testing (20%). The simulation is based on i9-9900K CPU with 64GB RAM. We train the models on the training set until they obtain the best performance on the validation set, and then test them on the testing set. All the models are trained with the same settings (i.e., batch size 16, Adam optimizer, Mean Squared Error (MSE) loss, batch normalization) to guarantee fairness.

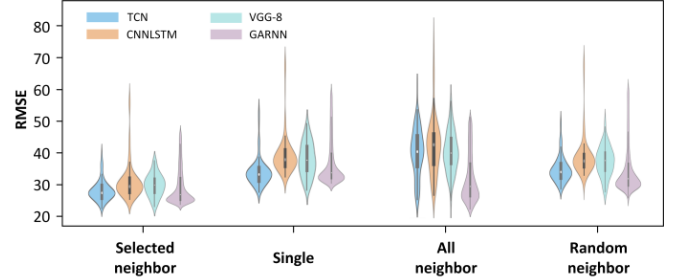

Fig. 9. Violin plots of 1-hour ahead CF forecasting RMSE on 95 PV sites.

TABLE VI
STATISTICS OF THE 1-HOUR AHEAD FORECASTING RMSE ON 95 PV SITES

| Scenarios | Evaluation Metrics | TCN | CNN-LSTM | VGG-8 | GARNN | Average |
|---|---|---|---|---|---|---|
| Selected neighbors | Media | 27.53 | 29.11 | 29.50 | 27.60 | **28.44** |
| | IQR | 4.78 | 5.65 | 4.77 | 8.41 | **5.90** |
| Single site | Media | 33.41 | 38.29 | 37.95 | 33.98 | 35.91 |
| | IQR | 3.92 | 6.84 | 9.08 | 7.22 | 6.77 |
| All sites | Media | 40.18 | 43.02 | 40.01 | 29.20 | 38.10 |
| | IQR | 11.77 | 10.55 | 8.56 | 10.59 | 10.37 |
| Random neighbors | Media | 34.05 | 37.88 | 37.91 | 31.96 | 35.45 |
| | IQR | 5.11 | 6.72 | 5.93 | 9.21 | 6.74 |

Violin plots of the 1-hour ahead forecasting RMSE on all the 95 PV sites are shown in Fig. 9 and Table VI. To validate the effectiveness of the neighbor selection algorithm, we train the 4 forecasting models using four kinds of inputs: 1) S*elected*

*Neighbors*: trained using historical data from selected neighbors. This is our target scenario. 2) *No Neighbors*: trained solely using historical data of the target PV site. 3) *All sites*: historical data of all PV sites are used for training without detector selection model. 4) *Random Neighbors*: detector sites are randomly selected.

From Fig. 9 and Table VI, we can see that
- When using measurements from only the selected neighbors, all models achieve the best performance (i.e. smallest RMSE median and IQR (Inter-Quantile Range)). This demonstrates the efficacy of the neighbor selection method.
- If measurements from all sites are used, the results have large RMSE variances, showing that data from uncorrelated sites can pollute the forecasting results.
- Using data from the target site or from randomly selected neighbors can lead to larger forecasting errors, showing the lack of spatial-temporal correlation information.

Next, the four deep-learning models with inputs from selected neighbors are compared for different forecasting horizons. In addition, to compare with non-machine learning methods, we include 2 other baseline methods: persistence model [36] and SARIMA [37]. From results as shown in Figs. 10, 11 and Table VII, we have the following observations:

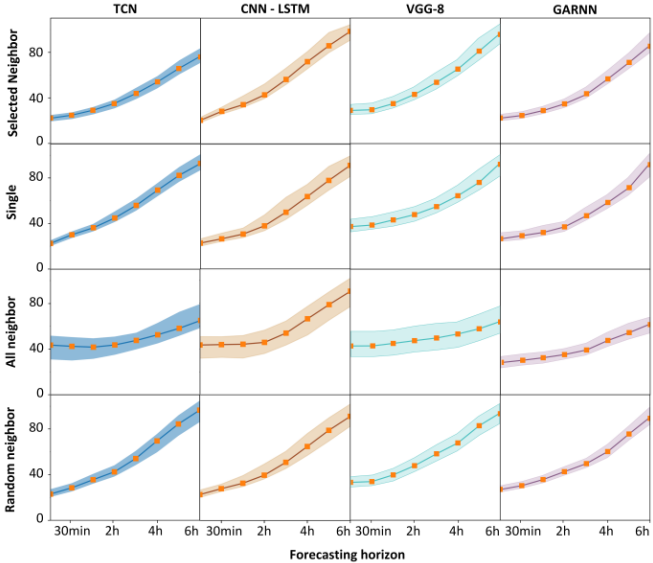

Fig. 10. CF forecasting RMSE for the 4 deep-learning models under different forecasting horizons with different detector selection strategies. The bandwidth represents the 90% confident interval (CI-90%) of the forecasting RMSE on 95 PV sites for measuring forecasting stability.

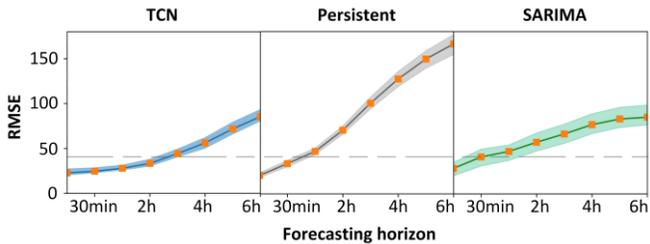

Fig. 11. Comparison of different forecasting models. The black horizontal dashed line shows the forecasting accuracy achieved in the TF by the physics-based model.

TABLE VII
FORECASTING PERFORMANCE EVALUATION (AVERAGED ON 95 SITES)

| Scenarios | Evaluation Metrics | TCN | CNN-LSTM | VGG-8 | GARNN |
|---|---|---|---|---|---|
| Selected neighbors | RMSE | **39.80** | 51.88 | 48.52 | 43.81 |
| | CI-90% | **10.37** | 15.81 | 16.25 | 10.89 |
| Single site | RMSE | **52.86** | 55.80 | 61.77 | 56.71 |
| | CI-90% | **11.67** | 18.00 | 15.03 | 12.66 |
| All sites | RMSE | 49.92 | 57.74 | 54.30 | **42.15** |
| | CI-90% | 17.84 | 23.33 | 25.69 | **14.52** |
| Random neighbors | RMSE | 54.60 | 52.26 | 58.11 | **49.77** |
| | CI-90% | 13.96 | 16.07 | 15.22 | **11.33** |
| Average computation time | | **≈ 6min** | ≈ 22min | ≈ 31min | ≈ 164min |

- TCN with detector sites achieves the lowest RMSE, best forecasting stability, and best computation efficiency.
- Using detector sites mainly improves the forecast accuracy of 2-3 hour-ahead PV forecast. Forecasting errors increase dramatically after hour 4 exceeding that of the physics-based model.
- We notice that The GARNN model achieves better performance than other methods if all sites are used or if detectors are randomly selected. This is because the attention mechanism in GARNN can dynamically identify the most correlated neighbors and assign them with higher weights, which acts as an "internal neighbor selection" mechanism. However, this also bring additional computing costs.

*C. TCN-based Forecast Reconciliation*

Data collected from May to September are used for training the reconciliation model. October data is used to test its performance. The TF forecasting horizon is 24 h and the CF is 6 h (i.e. the input/output length is 6×12 = 72 in Fig. 4) to reflect the most challenging CF forecasting scenario.

As shown in Fig. 12, the CF base forecasts use real-time data as inputs, it does not capture long-term trends well. Thus, the forecast curve starts to deviate from the actual when the forecasting horizon exceeds 2 hours. This phenomena can also be observed in Figs. 10 and 11. Thus, without inputs from NWP, CF suffers from large "trend errors". On the other hand, the NWP based TF forecast, while predicting the long-term trend well, cannot effectively capture intra-hour fluctuations. After reconciliation, the TF and CF forecasts can compensate each other. As the two forecasts are independently generated time series, the reconciled CF forecasts capture both intra-hour fluctuations and the long-term trend, achieving improved performance.

The reconciliation method is tested on all the 95 PV sites. As shown in Fig. 13 and Table VIII, after reconciliation, the forecasting RMSE is significantly reduced for forecasting horizon longer than 2 hours.

TABLE VIII
AVERAGE FORECASTING RMSE BEFORE AND AFTER RECONCILIATION

| Forecasting horizon | 5min | 30min | 2h | 4h | 6h | Average |
|---|---|---|---|---|---|---|
| Before reconciliation | **27.60** | **30.55** | 35.07 | 52.13 | 79.64 | 45.00 |
| After reconciliation | 28.30 | 30.68 | **32.71** | **36.84** | **38.07** | **33.22** |
| Improvement | -2.54% | -0.43% | 6.73% | 29.33% | 52.20% | 25.95% |

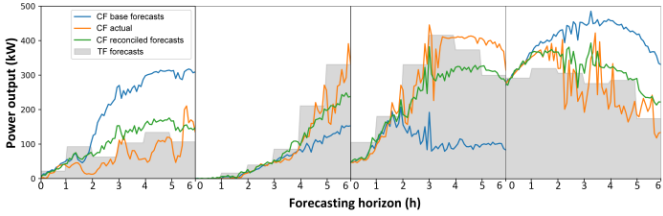

Fig. 12. Examples of CF forecasting results before and after reconciliation.

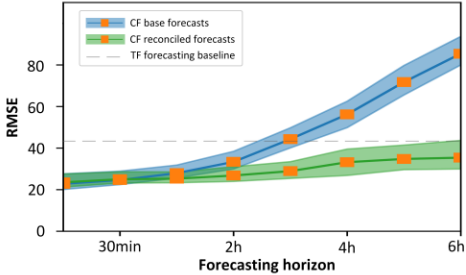

Fig. 13. CF Forecasting RMSE before and after reconciliation (95 PV sites).

## IV. CONCLUSION

This paper presents a TCN-based hybrid PV forecasting framework for utility-scale PV farms to improve the hours-ahead forecasting performance. In the TF model, TCN is used for NWP input data blending, which improves the forecasting accuracy by 37%. In the CF model, we develop a scenario-based detector sites selection algorithm so that real-time irradiance data from the most contributive neighbors of the target site can be used as inputs to the TCN-based hours-ahead PV forecast model. The TCN hours-ahead model uses spatial-temporal correlation between the detector and target sits to improve the forecast accuracy of large PV output drops at the target site. Then, we develop a forecast reconciliation process to better exploit the capability of the physics-based model for forecasting trend and the capability of the data-driven method for forecasting large cloud events. The results show that the proposed reconciliation model achieve another 26% performance improvement by merging the TF and CF forecasting results. We demonstrate that TCN is a comprehensive data-driven solution in solving the short-term PV forecasting problems with superior accuracy and computation efficiency.

## V. APPENDIX

### A. TCN

In this paper, TCN is used for NWP blending in the TF model, spatial-temporal forecasting in the CF model, and forecasting results reconciliation. The hyper-parameter settings for the 3 TCN models are given in the following Table IX.

The key principle of the TCN hyper-parameter selection is to guarantee the receptive field can cover the whole input sequences, according to (2). Besides, more filters mean stronger feature extraction ability. Skip connection is preferred when the network goes deeper to avoid gradient vanishing issue. We conducted an ablation study for the 3 key hyper-parameters that determine the receptive field, i.e., kernel size, dilation rate and number of stacks. Results show that the TCN model performance is stable and is insensitive to different hyper-parameter combinations as long as the receptive field is sufficient to cover the input sequences. However, the model performance will degrade when the receptive field is insufficient. Similar conclusions are also reported in [15].

TABLE IX
TCN MODEL CONFIGURATIONS

| parameters | NWP blending | CF forecasting | Forecasting reconciliation |
|---|---|---|---|
| Return sequence | True | False | True |
| Input data length | 48h | 6h | 6h |
| Kernel size | 3 | 2 | 3 |
| Number of filters | 32 | 64 | 32 |
| Dilation rate | [1, 2] | [0, 1, 3, 9] | [0, 1, 3, 9] |
| Number of stacks | 3 | 1 | 1 |
| Skip connection | Yes | No | No |
| **Model capacity** | ≈ 27K | ≈ 66K | ≈ 26K |

### B. VGG-8

VGG network is a very deep convolutional network proposed by Visual Geometry Group for image recognition purpose [33]. VGG is implemented to the spatial-temporal PV forecasting problem in [34] due to its outstanding feature extraction ability. In this paper we shrink the original 16-layer VGG network to an 8-layer VGG network (i.e. VGG-8) as a benchmarking method for TCN. Model configurations are given in the following Table X. The model capacity ≈ 61K.

TABLE X
VGG-8 CONFIGURATION

| Layer No. | Layer type | Filters /units | Kernel size | padding | activation |
|---|---|---|---|---|---|
| 1 | Conv2D | 8 | 3 | Same | ReLu |
| 2 | Conv2D | 8 | 3 | Same | ReLu |
| MaxPool2D (pool_size = 2, strides = 2, padding = same) | | | | | |
| 3 | Conv2D | 16 | 3 | Same | ReLu |
| 4 | Conv2D | 16 | 3 | Same | ReLu |
| MaxPool2D (pool_size = 2, strides = 2, padding = same) | | | | | |
| 5 | Conv2D | 32 | 3 | Same | ReLu |
| 6 | Conv2D | 32 | 3 | Valid | ReLu |
| MaxPool2D (pool_size = 2, strides = 2, padding = same) | | | | | |
| 7 | Dense | 128 | \ | \ | ReLu |
| 8 | Dense | 72 | \ | \ | Linear |

### C. CNN-LSTM

CNN-LSTM is another popular deep-learning model to solve the spatial-temporal PV forecasting problem [31]. The benchmarking CNN-LSTM model in this paper is given in the following Table XI with the model capacity ≈ 66K.

TABLE XI
CNN-LSTM CONFIGURATIONS

| CNN part | | LSTM part | | Dense part | |
|---|---|---|---|---|---|
| Kernel size | 7 | Number of units | 84 | Dense 1 | 128 units |
| Number of filters | 50 | Return sequence | False | Dense 2 | 72 units |
| Dropout rate | 0.2 | | | | |

### D. GARNN

GARNN model treats the PV sites as a graph, and the correlations among sites are represented by an adjacency matrix. This matrix can be time-varying because the correlation patterns among PV sites are determined by the dynamic weather conditions. To this end, a multi-head attention mechanism is introduced to learn a dynamic adjacency matrix from the time series data of each PV site. Then this dynamic matrix together with the original time series data will be fed into the RNN model to extract the temporal information and achieve forecasting. For more details please refer to [34][35]. The GARNN model configurations in this paper are given in the following Table XII. The model capacity $\approx$ 64K.

TABLE XII
GARNN CONFIGURATION

| Layer No. | Layer type | Key hyper-parameters |
|---|---|---|
| 1 | Attention | Embedding size = 16, Number of heads = 10 |
| 2 | GRU | Hidden features = 256, diffusions = 10 |
| 3 | GRU | Hidden features = 128, diffusions = 6 |
| 4 | Dense | Units = 192 |
| 5 | Dense | Units = 72 |